\documentclass[doublecol,comment]{epl2}
\usepackage{hyperref}
\usepackage{amsfonts}
\usepackage{amsthm}

\PassOptionsToPackage{pdfauthor={Vladimir V. Kisil},%
    pdftitle={Comment on `Do we have a consistent non-adiabatic quantum-classical mechanics?'},%
    pdfsubject={mathematics},%
    backref=page,%
  pdfkeywords={symmetries, quantum mechanic, quantum-classical bracket}
}{hyperref}

\providecommand{\Space}[3][]{\ensuremath{\mathbb{#2}^{#3}_{#1}{}}}
\providecommand{\intersect}[2]{\left.#1\right|_{#2}}
\providecommand{\eqref}[1]{\textup{(\ref{#1})}}
\providecommand{\ub}[3][]{\left\{\!#1\left[#2,#3\right]\!#1\right\}}
\providecommand{\anti}{\mathcal{A}}

\providecommand{\myhbar}{h}
\providecommand{\rmi}{\mathrm{i}}
\newtheorem{thm}{Theorem}
\theoremstyle{remark}
\newtheorem{example}[thm]{Example}
\providecommand{\revision}[1]{#1}

\title{Comment on `Do we have a consistent non-adiabatic
  quantum-classical mechanics?'}
\shorttitle{Comment on `Do we have a consistent\ldots'}

\author
{\href{http://www.maths.leeds.ac.uk/~kisilv/}{Vladimir V. Kisil}\thanks{On
    leave from Odessa University}, \email{\href{mailto:kisilv@maths.leeds.ac.uk}{kisilv@maths.leeds.ac.uk}}}
\shortauthor{Vladimir V. Kisil}

\institute{%
School of Mathematics,
University of Leeds,
Leeds LS2\,9JT,
UK.
Web: \href{http://maths.leeds.ac.uk/~kisilv/}%
{\textup{\texttt{http://maths.leeds.ac.uk/\~{}kisilv/}}}
}


\abstract{We argue with claims of the
  paper~\cite{AgostiniCapraraCiccotti07a} that the quantum-classic bracket
introduced in~\cite{Kisil05c} produces ``artificial coupling'' and has
``genuinely classical nature''.}

\providecommand{\Zbl}[1]{}
\begin{document}
\maketitle

\section{Introduction}
\label{sec:introduction}

This is a comment on the paper~\cite{AgostiniCapraraCiccotti07a}
, which evaluates the quantum-classical (QC) bracket:
\begin{eqnarray}
  {}[K_1,K_2]_{qc}&=& \frac{1}{ \rmi\myhbar} [K_1,K_2] 
  +\frac{1}{2}\left(\{K_1,K_2\}-\{K_2,K_1\}\right)
  \nonumber \\ && 
  -\intersect{\rmi\partial_{\myhbar_2} 
    [K_1,K_2]}{\myhbar_2=0}, \label{eq:qc-brackets}
\end{eqnarray}
introduced in paper~\cite[(26)]{Kisil05c}. The authors
in~\cite{AgostiniCapraraCiccotti07a} claimed that the QC
bracket~\eqref{eq:qc-brackets} exhibits:
\begin{itemize}
\item an artificial coupling property \revision{(i.e., coupling between the
  subsystems in the absence of an interaction)};
\item a genuinely classical nature \revision{(i.e., the apparent mixed quantum
  classical form reduces to a purely classical form for both
  subsystems)}. 
\end{itemize}

The assessment in~\cite{AgostiniCapraraCiccotti07a}
oversaw the following 
points:
\begin{enumerate}
\item QC bracket~\eqref{eq:qc-brackets} is the image of
  the universal bracket~\cite[(22)]{Kisil05c}:
  \begin{equation}\label{eq:p-brackets}
    \ub{k_1}{k_2}=(k_1*k_2-k_2*k_1) (\anti_1+\anti_2),
  \end{equation}
  under QC representation~\cite[(20)]{Kisil05c}. The universal bracket
  consists of convolution commutator and antiderivative
  operators~\cite[(12)]{Kisil05c}.  QC bracket requires consideration
  of the first jet space~\cite{Kisil05c}: \revision{the bracket is
  determined not only by their values of observables at
  \(\myhbar_2=0\) but also by the values of their first derivative
  with respect to \(\myhbar_2\) at zero (see the last term
  in~\eqref{eq:qc-brackets})}. 

\item The derivation QC bracket~\eqref{eq:qc-brackets} is independent 
  of p-me\-cha\-ni\-sa\-tion procedure introduced
  in~\cite[(23)]{Kisil05c}:
  \begin{eqnarray}
    \label{eq:mechanisation}
    q_j &\mapsto& Q_j=\delta'_{x_j}(g_1;g_2), \\
    \label{eq:mechanisation1}
    p_1 &\mapsto&
    P_j=\chi'_{s_k}(s_1+s_2)*\delta'_{y_j}(g_1;g_2), \quad
  \end{eqnarray}
  where \(j=1\), \(2\) and \(k=3-j\). 
\end{enumerate}
\revision{Here \emph{p-mechanisation}~\cite[\S~3.3]{Kisil02e}, as an
analog of quantisation, is a prescription how to build p-mechanical
observables out of classical ones.} It may not be very explicit
in~\cite{Kisil05c}, but the deduction of the
bracket~\eqref{eq:qc-brackets} is compatible with different choices
of p-mechanisation,\revision{ however the value of the bracket will be
different, see Exs.~\ref{ex:pure-class} and~\ref{ex:p-mechanosation}
below}.
To illustrate this in the present comment we use p-mechanisation given
by the Weyl (symmetric) calculus based on the following
correspondence, cf.~\cite[(23)]{Kisil05c},
\cite[(19)]{AgostiniCapraraCiccotti07a} and
\eqref{eq:mechanisation}--\eqref{eq:mechanisation1}:
\begin{equation}
  \label{eq:mechanisation-class}
  q_j \mapsto Q_j=\delta'_{x_j}(g_1;g_2), \quad
  p_j \mapsto
  P_j=\delta'_{y_j}(g_1;g_2), \quad
\end{equation}
Then the quantum-quantum image of the universal
bracket~\eqref{eq:p-brackets} of the respective coordinate and
momentum observables is:
\begin{equation}
  [Q_j,P_j]_{qq}=\frac{\myhbar_1+\myhbar_2}{\myhbar_k}I, \qquad k=3-j.
\end{equation}

Now we review the above two claims from the paper~\cite{AgostiniCapraraCiccotti07a}. 

\section{Artificial coupling property}
\label{sec:uncoupling-property}

There is the following claim
in~\cite[3001-p3]{AgostiniCapraraCiccotti07a}: ``It must be underlined
that eq.~(16) describes an artificial interaction even if the two
systems are not coupled by the
Hamiltonian.'' 
\revision{This coupling property is attributed to the fact that quantum-quantum
bracket in~\cite[(25)]{Kisil05c}
and~\cite[(16)]{AgostiniCapraraCiccotti07a} always contains both
Planck's constants \(\myhbar_1\) and \(\myhbar_2\), which are
generated by the presence of both antiderivative operators in the
definition of universal bracket~\eqref{eq:p-brackets}.}
 
\begin{example}
In order to exam the claim let us consider an uncoupled Hamiltonian
\(H(q_1,p_1,q_2,p_2)=H_1(q_1,p_1)+H_2(q_2,p_2)\). The p-mechanisation
(as well as quantisation) is a linear map~\cite[\S~3.3]{Kisil02e}, thus this uncoupled
structure will be preserved. Let \(\hat{B}\) be an observable
depending only from \(\hat{X}_2\) and \(\hat{D}_2\), thus it will
commute with \(H_1\). Therefore the commutator of \(B\) and \(H\) will
be the same as \(B\) and \(H_2\). The QC bracket is an image under a
representation of the usual commutator, thus the universal
bracket~\eqref{eq:p-brackets} of \(B\) and \(H\) will be the same as
\(B\) and \(H_2\). Consequently the \(\hat{H}_1\) will not affect the
dynamics of such an observable \(\hat{B}\).  
\end{example}
\revision{Therefore there is no coupling in the following meaning: arbitrary
change of the Hamiltonian \(H_1\) of the first subsystem will not
affect dynamics of any observable build from coordinates and momenta
of the second system only.}

\section{Genuinely classical nature}
\label{sec:genu-class-nature}

The paper~\cite[3001-p3]{AgostiniCapraraCiccotti07a} said ``In ref.
[8] it was suggested that the dynamical equation (16), in the limit
\(h_1 = h\) and \(h_2\rightarrow 0\), yields a QC dynamics.''
However the derivation in~\cite{Kisil05c} of the QC bracket
intentionally avoids any kind of semiclassic limits due to its
potential danger, see such an attempt in~\cite{Prezhdo-Kisil97} and
Example~\ref{ex:limit} below. The
actual method evaluates the image of the universal
bracket~\eqref{eq:p-brackets} under the QC
representation~\cite[(20)]{Kisil05c} of the group \(\mathbb{D}^m\).

The paper~\cite[3001-p4]{AgostiniCapraraCiccotti07a} ``corrected'' the
original derivation of QC bracket replacing the initial set of Planck
constants \(\myhbar_1\) and to \(\myhbar_2\) by the new one
\(\myhbar_{\mathrm{eff}}\) defined by the expression:
\begin{equation}
  \frac{1}{\myhbar_{\mathrm{eff}}}=\frac{1}{\myhbar_1} + \frac{1}{\myhbar_2}.
\end{equation}
However this transformation is singular for \(\myhbar_1\myhbar_2=0\)
and needs special clarifications how to proceed for such values.  
\begin{example}
\label{ex:limit}
  Let us consider the transformation \(U_h: f(x,y) \mapsto
  f(hx,\frac{1}{h}y)\), which is a unitary operator \(L_2(\mathbb{R}^2)
  \rightarrow L_2(\mathbb{R}^2)\) for any \(h>0\). However this does not
  allow us ``to take the limit \(h\rightarrow 0\)'' through the
  straightforward substitution \(h= 0\).
\end{example}

Furthermore the paper~\cite[3001-p3]{AgostiniCapraraCiccotti07a}
claims that  ``we have shown that the 
equation of motion (16) does not lead to a non-trivial 
QC limit''. \revision{However, this is caused by
p-mechanisation~\eqref{eq:mechanisation}--\eqref{eq:mechanisation1},
cf. the next two examples. }
\begin{example}
  \label{ex:pure-class}
\revision{  Let \(B_1\) and \(B_2\) are squares of coordinate \(Q\) and momentum
  \(P\)
  observables (of the quantum subsystem) respectively.  Under
  p-mechanisation~\eqref{eq:mechanisation}--\eqref{eq:mechanisation1}
  they are represented by squares of corresponding
  convolutions
  . Then the commutator (first term of bracket~\eqref{eq:qc-brackets})
  of their QC representations is zero, the second term
  in~\eqref{eq:qc-brackets}) vanishes since no classical observables
  present, and the third termin~\eqref{eq:qc-brackets}) is equal to QC image of the observable
  \(4QP\). Thus the total bracket is indeed the same as the Poisson
  bracket for those observables.}
\end{example}


Let us examine the above claim for the
p-mechanisation~\eqref{eq:mechanisation-class} and assume that two
p-mechanical observables \(B_1\) and \(B_2\), that is two convolutions
on the group \(\Space{D}{n}\)~\cite[p.~876]{Kisil05c}, for any fixed
\(g_1\) are multiples of the delta function in \(g_2\), e.g. as in
Ex.~\ref{ex:pure-class}.  Under the QC representation
\(\rho_{(h;q,p)}\)~\cite[(20)]{Kisil05c} those observables become
operators \(\rho_{(h;q,p)}(B_1)\) and \(\rho_{(h;q,p)}(B_2)\) on the
state space for the quantum subsystem without any dependence from
classical coordinates \(p\), \(q\) and the respective Planck constant
\(\myhbar_2\). Correspondingly the second and the third terms of the
bracket~\eqref{eq:qc-brackets} vanish and this bracket is equal to the
(quantum) commutator \(\frac{1}{\rmi \myhbar}[\rho_{(h;q,p)}(B_1),
\rho_{(h;q,p)}(B_2)]\).

Therefore if we admit the
claim~\cite[3001-p3]{AgostiniCapraraCiccotti07a} that
QC bracket~\eqref{eq:qc-brackets} always coincides with
the purely classic Poisson bracket, then we have to accept that any
quantum commutator is always equal to the Poisson bracket.
\begin{example}
  \label{ex:p-mechanosation}
  Under p-mechanisation~\eqref{eq:mechanisation-class} the squares of
  momentum and coordinates from Ex.~\ref{ex:pure-class} are
  represented by convolutions with kernels
  \(\delta''_{x_1x_1}(g_1;g_2)\) and \(\delta''_{y_1y_1}(g_1;g_2)\).
  Their commutator on the group \(\Space{D}{1}\) is
  \(4\delta'''_{x_1y_1s_1}+2\delta''_{s_1s_1}\).  Thus the universal
  bracket~\eqref{eq:p-brackets} is
  \begin{equation}
    \ub{B_1}{B_2}=4\delta''_{x_1y_1}+2\delta'_{s_1}+(4\delta'''_{x_1y_1s_1}+2\delta''_{s_1s_1})\anti_2.
  \end{equation}
  In the QC representation of \(\Space{D}{1}\) the last term of the
  sum vanishes and two first terms produce \(4QP+2\rmi\myhbar I\).
  This is the quantum commutator of \(Q^2\) and \(P^2\) times
  \(\frac{1}{\rmi\myhbar}\). There is no unitary representation to
  get rid of the purely imaginary term \(2\rmi\myhbar I\) in order
  to reduce the QC bracket of \(B_1\) and \(B_2\) to the value
  \(4QP\) of their Poisson bracket.
\end{example}

\section{Conclusion}
\label{sec:conclusion}
In this paper we demonstrated that the QC
bracket~\eqref{eq:qc-brackets} does not possess itself two properties
of ``artificial coupling'' and ``genuinely classical nature'' as
claimed in~\cite{AgostiniCapraraCiccotti07a}.
Unfortunately the claims~\cite{AgostiniCapraraCiccotti07a} were
uncritically translated by some other authors,
see~\cite{Hall08a,ZhangLinWu08a}
.

\revision{We showed that for a decoupled Hamiltonian the dynamics of observables
localised in one subsystem is unaffected by the Hamiltonian of the
other subsystem. The ``classical nature'' described
in~\cite{AgostiniCapraraCiccotti07a} is rooted in the p-mechanisation
used in~\cite{Kisil05c} and does not appear with other choice of
p-mechanical observables.}  

The main conclusion of the commented
paper~\cite{AgostiniCapraraCiccotti07a} is%
: ``%
We suggest that a different Ansatz
for the equations of motion, could indeed produce non-trivial
QC equations''.  This comment is aimed to clarify
possible directions for such a search.

\section{Acknowledgements}
\label{sec:acknowledgments} I am grateful to Dr.~Frederica Agostini
for useful discussions and providing me with a part of her unpublished
thesis
. Prof.~O.V.~Prezhdo \revision{and an anonymous referee}
made suggestions, which improved presentation in this comment.

\small

\begin{thebibliography}{1}
\expandafter\ifx\csname url\endcsname\relax\def\url#1{\texttt{#1}}\fi

\bibitem{AgostiniCapraraCiccotti07a}
\Name{Agostini F., Caprara S. \and Ciccotti G.} \REVIEW{Europhys. Lett. EPL
  }{78}{2007}{Art. 30001, 6} \doi{10.1209/0295-5075/78/30001}.


\bibitem{Prezhdo-Kisil97}
\Name{Prezhdo O.~V. \and Kisil V.~V.} \REVIEW{Phys. Rev. A (3) }{56}{1997}{162}
  \arXiv{quant-ph/9610016}.


\bibitem{Kisil02e}
\Name{Kisil V.~V.} \REVIEW{J. Phys. A }{37}{2004}{183}
  \arXiv{quant-ph/0212101},
  \href{http://stacks.iop.org/0305-4470/37/183}{On-line}. \Zbl{1045.81032}.

\bibitem{Kisil05c}
\Name{Kisil V.~V.} \REVIEW{Europhys. Lett. }{72}{2005}{873}
  \arXiv{quant-ph/0506122},
  \href{http://dx.doi.org/10.1209/epl/i2005-10324-7}{On-line}.

\bibitem{Hall08a}
\Name{Hall M. J.~W.} \REVIEW{Physical Review A}{78}{2008}{042104}.

\bibitem{ZhangLinWu08a}
\Name{Zhan F., Lin Y. \and Wu B.} \REVIEW{Journal of Chemical Physics}
  {128}{2008}{315204}.


\end{thebibliography}
\newcommand{\noopsort}[1]{} \newcommand{\printfirst}[2]{#1}
  \newcommand{\singleletter}[1]{#1} \newcommand{\switchargs}[2]{#2#1}
  \newcommand{\irm}{\textup{I}} \newcommand{\iirm}{\textup{II}}
  \newcommand{\vrm}{\textup{V}} \providecommand{\cprime}{'}
  \providecommand{\eprint}[2]{\texttt{#2}}
  \providecommand{\myeprint}[2]{\texttt{#2}}
  \providecommand{\arXiv}[1]{\myeprint{http://arXiv.org/abs/#1}{arXiv:#1}}
  \providecommand{\CPP}{\texttt{C++}} \providecommand{\NoWEB}{\texttt{noweb}}
  \providecommand{\MetaPost}{\texttt{Meta}\-\texttt{Post}}
  \providecommand{\GiNaC}{\textsf{GiNaC}}
  \providecommand{\pyGiNaC}{\textsf{pyGiNaC}}
  \providecommand{\Asymptote}{\texttt{Asymptote}}

\end{document}